\documentclass[twocolumn,showpacs,preprintnumbers,amsmath,amssymb,superscriptaddress,prl]{revtex4}
\pdfoutput=1
\usepackage{graphicx}
\usepackage{dcolumn}
\usepackage{bm}
\usepackage{epsfig}
\usepackage[american]{babel}
\usepackage{color}
\usepackage{textcomp}
\usepackage{amsmath}
\usepackage{amssymb}
\usepackage{amsfonts}
\usepackage[hang,normalsize,sf,SF]{subfigure}

\newcommand{\un}[1]{\,\mathrm{#1}}

\begin{document}

\title{Telegraph Noise in Coupled Quantum Dot Circuits Induced by a Quantum Point Contact}

\author{D.\ Taubert}

\affiliation{Center for NanoScience and Fakult\"at f\"ur Physik, Ludwig-Maximilians-Universit\"at, Geschwister-Scholl-Platz 1, D-80539 M\"unchen, Germany}

\author{M.\ Pioro-Ladri\`{e}re}

\affiliation{Quantum Spin Information Project, ICORP, Japan Science and Technology Agency, Atsugi-shi, 243-0198, Japan}

\author{D.\ Schr\"oer}

\affiliation{Center for NanoScience and Fakult\"at f\"ur Physik, Ludwig-Maximilians-Universit\"at, Geschwister-Scholl-Platz 1, D-80539 M\"unchen, Germany}

\author{D.\ Harbusch}

\affiliation{Center for NanoScience and Fakult\"at f\"ur Physik, Ludwig-Maximilians-Universit\"at, Geschwister-Scholl-Platz 1, D-80539 M\"unchen, Germany}

\author{A.\,S.\ Sachrajda}

\affiliation{Institute For Microstructural Sciences, NRC, Ottawa, K1A 0R6, Canada}

\author{S.\ Ludwig}

\affiliation{Center for NanoScience and Fakult\"at f\"ur Physik, Ludwig-Maximilians-Universit\"at, Geschwister-Scholl-Platz 1, D-80539 M\"unchen, Germany}

\begin{abstract}
Charge detection utilizing a highly biased quantum point contact has become the most effective probe for studying few electron quantum dot circuits. Measurements on double and triple quantum dot circuits is performed to clarify a back action role of charge sensing on the confined electrons. The quantum point contact triggers inelastic transitions, which occur quite generally. Under specific device and measurement conditions these transitions manifest themselves as bounded regimes of telegraph noise within a stability diagram. A nonequilibrium transition from artificial atomic to molecular behavior is identified. Consequences for quantum information applications are discussed.
\end{abstract}


\maketitle

The experimental demonstration of single spin isolation by purely electrostatic means \cite{Tarucha96,Ciorga00} has spurred research into the application of quantum dots (QDs) as spin qubits utilizing the exchange interaction \cite{Loss98}. This has led to the development of multiple few electron QD circuits and proof of concept demonstrations of the DiVincenzo criteria of quantum computing \cite{Elze04,Kopp04,Nowa07,Pett07}. Investigated circuits have already advanced to few electron triple QD devices \cite{Gaud06,Gaud07,Schr07,Rogg07,Amaha07}. For more than one QD connected in series to source and drain reservoirs, electron transport becomes severely limited as a detection tool. Transfer of an electron from one lead to the other by sequential tunneling between QDs requires the very restrictive condition of matched chemical potentials of the leads and all QDs; a condition that only occurs at singular points in the stability diagram. Instead a capacitively coupled quantum point contact (QPC) is often utilized as a detector sensitive to the charge configuration of the multi-qubit circuit \cite{Gaud06,Schr07}. For progress to be made in quantum information applications using these qubits, it is very important to understand the level to which this charge detection technique is  noninvasive.

In this Letter we present results from various double and triple QD circuits where we observe a noise of telegraphic nature \cite{Gust07} but located in well--defined regions of the stability diagram and under precise lead coupling conditions. We discuss the origin of the noise and confirm that it is related to a form of back action from the QPC being used to monitor the charge state of the coupled QDs. The back action enables an otherwise forbidden interchange of charge between the QDs and the leads \cite{Onac06,Khra06}. An important conclusion from this Letter is that the noise can be observed for specific lead couplings in bounded regions in a stability diagram, but there is no known reason for the back action not to extend beyond these boundaries. We will discuss the consequences of this effect for spin--qubit measurements.

The measurements presented here are performed at electron temperatures of $30\un{mK}\lesssim T\lesssim 100\un{mK}$ on three different GaAs/AlGaAs heterostructures each containing a two dimensional electron gas between 90\,nm and 120\,nm beneath the surface. Charge carrier densities and mobilities vary between $1.7\times 10^{15}\un{m^{-2}}\lesssim n_{\rm s}\lesssim 2.8\times 10^{15}\un{m^{-2}}$ and $75\un{m^2/Vs}\lesssim\mu\lesssim 200\un{m^2/Vs}$. Metal gates defined on the surface by electron beam lithography are negatively biased in order to laterally define the nanoscale devices.

\begin{figure}[ht]
\includegraphics[clip,width=.435\textwidth]{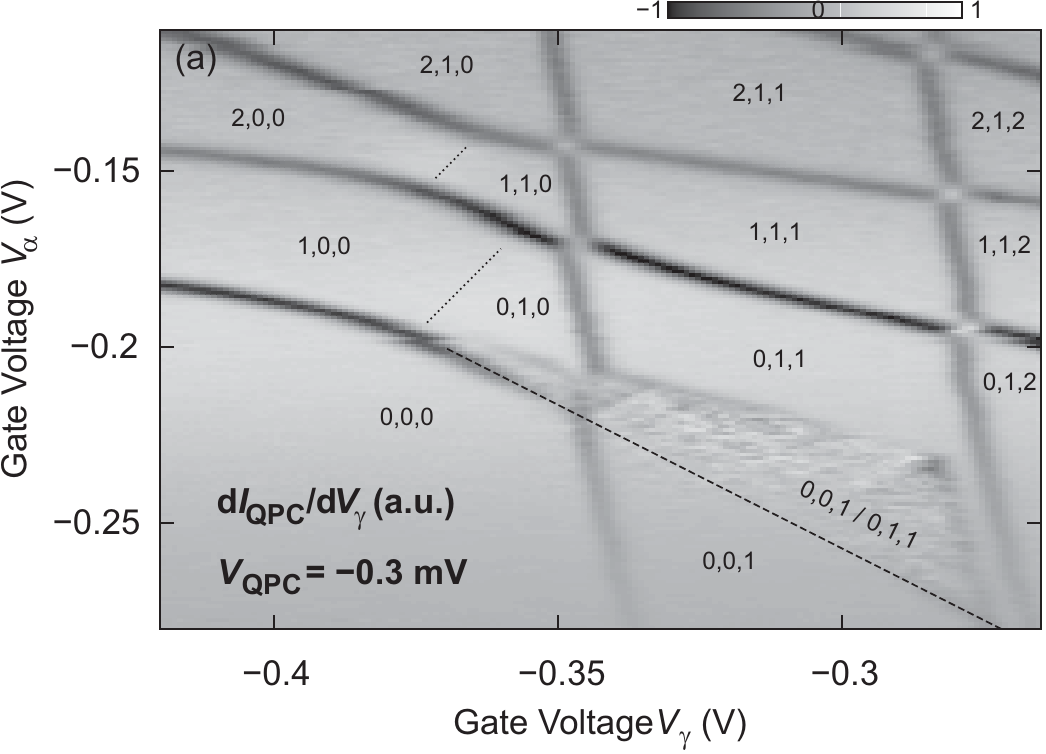}
\includegraphics[clip,width=.435\textwidth]{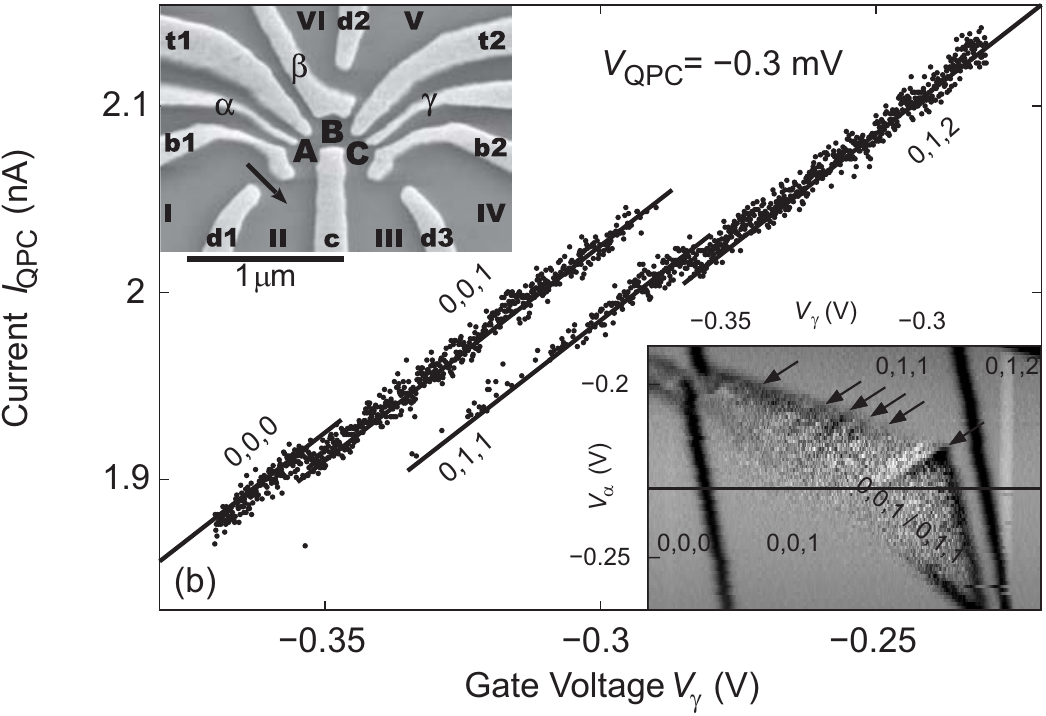}
\vskip-2mm
\caption{(a) Stability diagram of a triple QD defined by gates shown in (b). Grayscale plot of transconductance of the lhs QPC (arrow). Dashed (dotted) lines indicate a charging (reconfiguration) line(s) involving QD $B$. Occupied configurations are labeled with the number of electrons $N_A,N_B,N_C$ charging the QDs. (b) dc--current $I_{\rm QPC}$ at $V_\alpha=-0.23\un{V}$ along the horizontal line in the rhs inset. Straight lines mark $I_{\rm QPC}$ expected for specific charge configurations. Lhs inset: Scanning electron micrograph of the sample surface (gates in light gray). Positions of QDs $A$, $B$, and $C$, the current path (arrow) and ohmic contacts (roman numbers) are also indicated. Rhs inset: Transconductance as in (a) but at slightly weaker tunnel couplings to the leads.}
\vskip-2mm
\label{fig1}
\end{figure}

The inset in Fig.~\ref{fig1}(b)
shows a gate geometry capable to define a serial triple QD and up to three QPCs \cite{Schr07,Schr_PhD}. Only the left--hand side (lhs) QPC is used as charge detector for the three QDs $A$, $B$, and $C$ [Fig.~\ref{fig1}(b)]. Ohmic contact I is biased at $V_{\rm QPC}=-0.3\,\un{mV}$ against all other ohmic contacts kept at identical potentials. Figure \ref{fig1}(a) shows a two-dimensional charge stability diagram plotting the transconductance ${\rm d} I_{\rm QPC}/{\rm d} V_\gamma$ of the lhs QPC as a function of plunger gate voltages $V_\alpha$ and $V_\gamma$, while $V_\gamma$ is modulated with $0.66\un{mV}$ at a frequency of $33\un{Hz}$ (integration time is $0.3\un{s}$). Transconductance minima with three different main slopes mark charging events of the three QDs \cite{Schr07}. These charging lines and the charge reconfiguration lines connecting pairs of triple points (transconductance maxima) separate occupied configurations $N_A,N_B,N_C$ of the triple QD, where $N_{A,B,C}$ electrons charge each QD.

A very exceptional feature in the stability diagram is a triangular shaped area of strong telegraph noise. It is plotted again in the right--hand side (rhs) inset of Fig.~\ref{fig1}(b) for weaker tunnel couplings of the triple QD to its leads. The dashed line in Fig.~\ref{fig1}(a) indicates the expected position of the first charging line of the central QD (B). Here, the ground state configuration changes from $0,0,1$ to $0,1,1$. Figure \ref{fig1}(b) plots the dc--current $I_{\rm QPC}$ independently measured along the horizontal line in the rhs inset with the current--amplifier set to an integration time of $0.3\,\un{ms}$. The overall positive slope of $I_{\rm QPC}(V_\gamma)$ reflects the capacitive coupling of the QPC to gate $\gamma$. Straight lines are added to approximate the current values expected at four incidental charge configurations. Their distances are related to the capacitive coupling differences between each QD and the QPC. Transitions of the current between adjacent lines are observed at charging lines in the stability diagram. Within the triangle of strong telegraph noise $I_{\rm QPC}$ fluctuates between two lines, suggesting a bistability of the triple QD jumping between two configurations, namely $0,0,1$ and $0,1,1$. These fluctuations are equivalent to an electron tunneling into and out of the central QD. The excited state 0,0,1 is thereby significantly occupied (even more than the ground state configuration 0,1,1), suggesting a non--equilibrium situation. Rectification measurements on a single QD place a limit of $20\un{\mu eV}$ on the magnitude of external noise. A possible non--equilibrium energy source is the QPC used for charge detection.

The observation of such novel noise patterns is directly related to the rates for electron tunneling in or out of the ground state configuration, $\Gamma_{\rm in,out}$. For $\Gamma_{\rm in}=\Gamma_{\rm out}$ the stability diagram features charging lines which in thermal equilibrium separate ground state configurations. Charge fluctuations can be resolved in transconductance measurements [Fig.~\ref{fig1}(a)] if the spectra of both $\Gamma_{\rm in}$ and $\Gamma_{\rm out}$ overlap with the modulation frequency, and in dc--measurements [Fig.~\ref{fig1}(b)] if $\Gamma_{\rm in,out}^{-1}$ are longer than the integration time \cite{Gust07}. Small rates $\Gamma_{\rm in,out}$ naturally occur along charging lines of the central QD of a serial triple QD. Here, intermediate excited states (in our case involving configurations $1,0,1$ or $0,0,2$) will be occupied while the charge of the central QD fluctuates, thus, making these two--step processes ($0,1,1 \leftrightarrow 0,0,1$) slow. The intermediate states can be excited by absorption of energy or virtually occupied in a cotunneling process.

\begin{figure}[ht]
\vskip-2.5mm
\centering\includegraphics[clip,width=.435\textwidth]{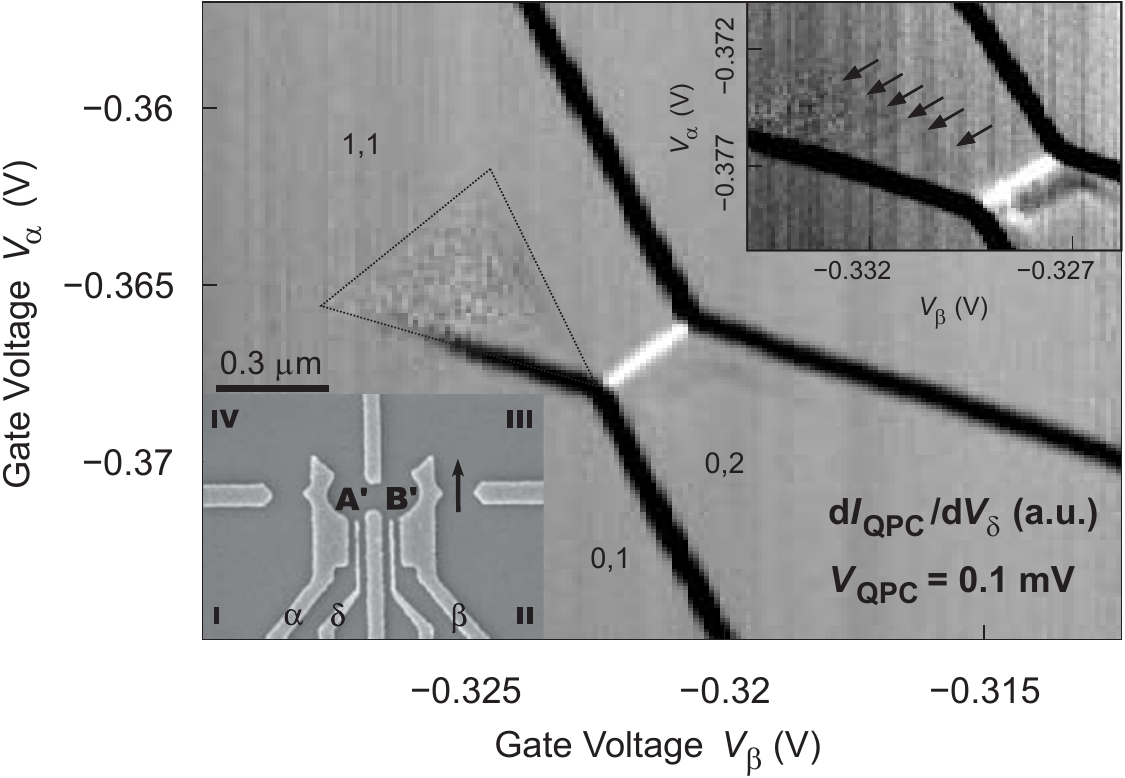}
\vskip-2mm
\caption{Stability diagram of a double QD defined by the gates shown in lhs inset. The grayscale plots the transconductance of the rhs QPC (arrow) at $V_{\rm{II}}\equiv V_{\rm QPC}=-0.1\un{mV}$ as a function of the gate voltages $V_\alpha$ and $V_\beta$, while $V_\delta$ is modulated with $1\un{mV}$ at a frequency of $23\un{Hz}$ using an integration time of $0.3\un{s}$. Occupied configurations are labeled with the number of electrons $N_{ A'},N_{ B'}$ charging the QDs. Dotted lines enclose a region of telegraph noise. Rhs inset: Similar measurement at stronger inter--dot tunnel coupling. Lhs inset: Scanning electron micrograph showing the gate layout. Positions of the QDs $A'$ and $B'$, the current path and ohmic contacts are indicated.}
\label{fig2}
\vskip-2mm
\end{figure}

If one of the peripheral QDs $A$ or $C$ of the triple QD is replaced by a large tunnel barrier, an equivalent situation can be achieved in a double QD. Figure \ref{fig2}
shows a charge stability diagram of a tunnel coupled double QD, detected with a nearby QPC. Here, the lhs QD $A'$ is tuned to be  coupled very weakly to the lhs lead while all other tunnel couplings are much larger. Such an arrangement makes the charging of QD  $A'$ difficult, i.\,e.\ when moving in Fig.~\ref{fig2} across the chosen configurations $0,1 \leftrightarrow 1,1$ or equivalent transitions $N_{A'},N_{B'} \leftrightarrow N_{A'}+1,N_{B'}$. For the triple QD, charging the central QD involves an intermediate excited state, i.\,e.\ $1,0,1$ or $0,0,2$. For a double QD one of these two--step processes is replaced by one direct but very slow tunneling process. As the tunneling rate $0,1 \leftrightarrow 1,1$ charging QD $A'$ becomes smaller than the modulation frequency, the associated charging line completely vanishes. Hence direct tunneling events involving the lhs lead become increasingly infrequent. In this regime, it is possible that charging QD $A'$ via QD $B'$ becomes the faster process. This situation is very similar to that in a triple QD at the charging line of the central QD. In both cases charging events involve an intermediate excited state.

Above the disappearing charging line a triangular shaped region of strong telegraph noise (marked by dotted lines) is observed. Here, the double QD fluctuates between the ground state configuration $1,1$ and the configuration $0,1$. Along the extension of another charging line between $0,1$ and $0,2$ the fluctuations stop. Below this extension line the rates $\Gamma_{\rm in}$ and $\Gamma_{\rm out}$ are small because the occupation of the intermediate state $0,2$ requires energy. Above this extension line $\Gamma_{\rm in}$ rapidly increases, because here the transition from $0,1$ via $0,2$ to the ground state configuration $1,1$ is possible by sequential tunneling without absorption of energy. On the other hand $\Gamma_{\rm out}$ stays small because starting at the ground state, energy needs to be absorbed always. Hence, above the extension line the ground state is mostly occupied. The bistable triangle of the triple QD (Fig.~\ref{fig1}) shows an analogous behavior; the two higher edges are extensions of charging lines of QDs $A$ (more horizontal line) and $C$ (more vertical line). Above or to the right of these two edges (starting from the excited configuration $0,0,1$) the intermediate configurations $1,0,1$ or $0,0,2$ and thereafter the ground state ($0,1,1$) can be occupied without energy absorption. These situations allow an electron to rapidly tunnel sequentially into the central QD, while $\Gamma_{\rm out}$ stays slow. Then the ground state is mostly occupied.

Across the third line of the bistable triangle of the double QD the telegraph noise ceases gradually together with the charging line. This smeared transition line is approximately parallel to the charge reconfiguration line (white) between triple points. Therefore, the detuning between configurations $1,1$ and $0,2$ is constant. The disappearance of telegraph noise across this line implies that here $\Gamma_{\rm in}$ and $\Gamma_{\rm out}$ mainly depend on the detuning between the two QDs. Hence, co--tunneling between configurations $1,1$ and $0,1$ is slower than processes in which the intermediate configuration $0,2$ is actually occupied. The necessary energy can be provided by the nearby QPC.

The rhs inset in Fig.~\ref{fig2} shows the bistable region for somewhat stronger coupling to the rhs lead but otherwise similar conditions. This region is patterned by several lines (arrows in inset of Fig.~\ref{fig2}) with constant detuning between the two QDs (parallel to the charge reconfiguration line). A similar pattern is observed for the triple QD [arrows in Fig.~\ref{fig1}(b)]. Here, one of these lines starts at the tip of the triangle. Along this line the electronic ground states of the intermediate configurations $1,0,1$ and $0,0,2$ are degenerate, enhancing tunneling between them and thereby decreasing occupation of the ground state configuration $0,1,1$. Other lines (arrows) indicate resonances of excited electronic states of the same intermediate configurations. These excited states can be occupied by absorption of energy quanta provided by the driven QPC. Note that the distances between lines are not determined by single QD excitation energies but can be much smaller when exited states of QDs with unequal excitation spectra are aligned.

For the double QD the parallel lines (rhs inset Fig.~\ref{fig2}) indicate a resonance between electronic states of the configurations $1,1$ and $0,2$. As described above for the case of the triple QD, the resonance condition effects the ground state occupation. Dark and bright lines below the charge reconfiguration line of the double QD (Fig.~\ref{fig2}) are also believed to be due to related phenomena.

\begin{figure}[th]
\vskip-2mm
\centering{\includegraphics[clip,width=.435\textwidth]{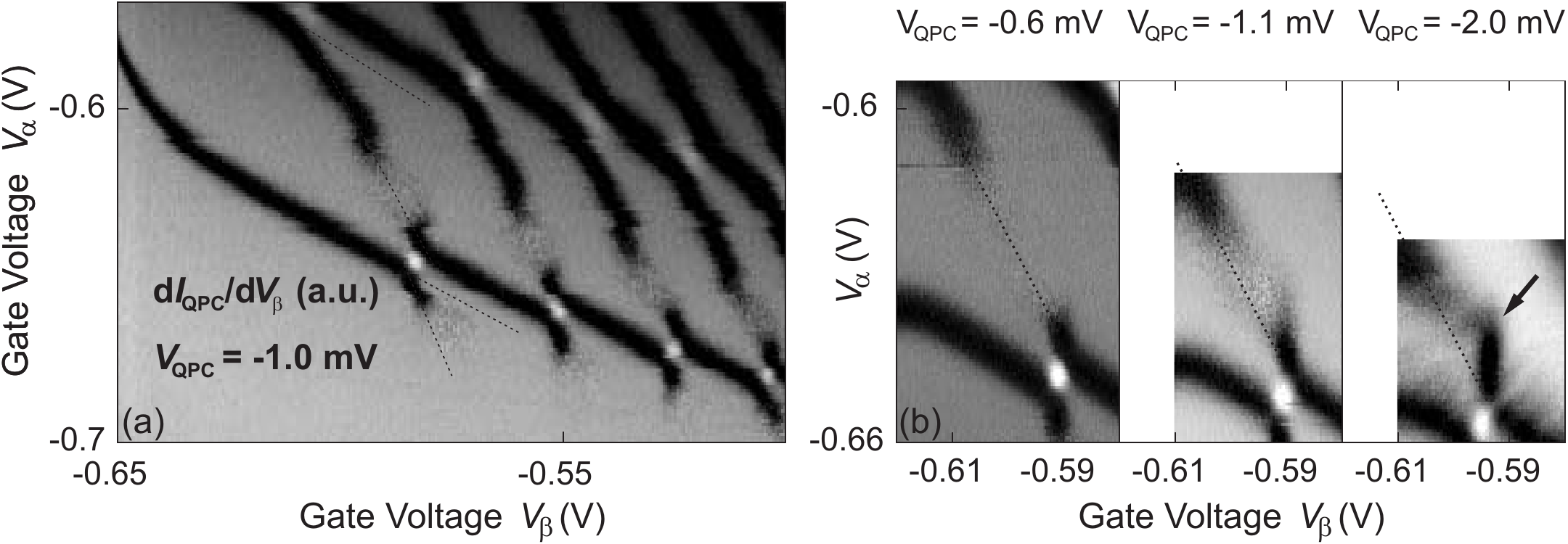}}
\vskip-2mm
\caption{(a) Stability diagram of a double QD detected with the lhs QPC with $V_\beta$ modulated by $1\un{mV}$ at a frequency of $377\un{Hz}$. The gate layout is identical to that shown in Fig.~\ref{fig1} but on a different wafer. QD $C$ is not defined ($V_\gamma=V_{\rm b2}=0$) and QD $B$ is very weakly tunnel coupled to the rhs lead. (b) Blow-up of a gapped charging line from (a) for three different bias voltages applied across the lhs QPC. Dotted lines are explained in the main text.}
\label{fig3}
\vskip-3mm
\end{figure}

In general, triangular regions of telegraph noise can be observed in a stability diagram of coupled QDs, as long as throughout the bistable region the spectrum of the telegraph noise overlaps with the modulation frequency of the transconductance measurement. However, where the modulation frequency exceeds the telegraph noise spectrum, which varies throughout the stability diagram, no fluctuations will be observed. Figure \ref{fig3}(a)
displays a stability diagram of a double QD (QDs $A$ and $B$) in the few electron regime defined on a different wafer but with an identical gate layout as shown in Fig.~\ref{fig1} \cite{Schr_PhD}. QD $B$ is tuned to be coupled very weakly to its lead. The transconductance ${\rm d}I_{\rm QPC}/{\rm d}V_\beta$ of the lhs QPC is measured with $V_{\rm QPC}=-1.0\un{mV}$ at a relatively high frequency of $377\un{Hz}$.

As $V_\alpha$ is decreased each charging line of QD $B$ (steeper lines) is interrupted, reappears at triple points, and then completely disappears. Fluctuations in these regions suggest that the rates for charging QD $B$ $\Gamma_{\rm in}\simeq\Gamma_{\rm out}$ are comparable to the modulation frequency of $V_\beta$. Before the lines finally vanish, the onsets of triangular regions of fluctuations can be seen. Dotted lines in Fig.~\ref{fig3}(a) mark boundaries of expected bistable regions. Here they are partly observed, i.\,e.\ wherever rates $\Gamma_{\rm in,out}$ match the modulation frequency. The transient reappearance of charging lines at triple points is a consequence of electrons occupying delocalized molecular states of both QDs at triple points enhancing $\Gamma_{\rm in}$.

Fig.~\ref{fig3}(b) plots an interrupted charging line for three different $V_{\rm QPC}$. For the smallest $|V_{\rm QPC}|=0.6\un{mV}$ the line is merely interrupted. For increasing $|V_{\rm QPC}|$ the shape of the charging line changes while the gap decreases (the parallel dotted lines mark the expected position of the charging line \cite{comm-1}). Since in thermal equilibrium the position of a charging line marks a degeneracy of two ground state configurations, such an externally induced change of charging line shapes can only be explained within a non--equilibrium situation. Here the external parameter is $V_{\rm QPC}$ which strongly suggests that the QPC acts as non--equilibrium energy source.

The decreasing size of the gap in the charging line (on which $\Gamma_{\rm in}=\Gamma_{\rm out}$) suggests that both rates grow with $|V_{\rm QPC}|$. The deformation towards more positive gate voltages with growing $|V_{\rm QPC}|$ implies a stronger increase of $\Gamma_{\rm out}$ compared to $\Gamma_{\rm in}$. A kink develops at large $|V_{\rm QPC}|$ [arrow in Fig.~\ref{fig3}(b)] and divides the charging line into two parts with distinct slopes, suggesting a competition between two channels for $\Gamma_{\rm in}$. For the sharper vertical line, $\Gamma_{\rm in}$ is dominated by direct tunneling from the lhs lead into the rhs QD $B$ via a molecular state delocalized between both QDs. Along the flatter part of the charging line these direct processes are slow compared to charging via an intermediate excited state with an electron occupying QD $A$. $\Gamma_{\rm out}$ is always determined by the indirect process, since starting from the ground state, energy must be absorbed before an electron can escape QD $B$.

Charge fluctuations will generally occur in QDs as long as a noise source out of thermal equilibrium, such as a biased QPC, provides enough energy. Such fluctuations are seldom resolved in stability diagrams \cite{Gust07} because typical detection schemes are too slow. In recent coherent experiments \cite{Pett07,Avin04,Buks98} a QPC used as charge detector is constantly biased at a few hundred microvolts. Then the coherence time will be influenced by the back action of the QPC \cite{comment}. The non--equilibrium nature of the fluctuations we observed indicates that this back action goes beyond the usually considered quantum mechanical entanglement of the detector and a coherent state \cite{Avin04,Buks98}. The important implication for spin--qubit experiments is that the detector should be switched off during coherent operation. It is fair to assume that the QPC also causes energy relaxation processes between electronic states within a single
charge configuration even though they are not resolved by charge detection experiments. Nevertheless such processes will also cause decoherence.

In conclusion we have observed and explained a novel form of telegraph noise that is triggered by the backaction of a QPC charge detector. The frequency of the fluctuations strongly depends on the resonant tunnel coupling between the ground state and other charge configurations. We stress that while in double QDs it is necessary to configure the device in a specific way to observe the bistability in a low frequency measurement, the necessary weak coupling to the leads condition arises automatically for a serial triple QD device. This will also be the case for quantum circuits containing an even larger number of QDs. Our results do not distinguish between fluctuations caused by photon versus phonon assisted tunneling, which needs to be clarified in future experiments.

The authors thank F.\ Marquardt, S.\ Kehrein, J.\,P.\ Kotthaus, S.\ Studenikin and P.\ Hawrylak, for fruitful discussions, K.\ Eberl (Fig.~\ref{fig1}), Z.\,R.\ Wasilewski, J.\,A.\ Gupta (Fig.~\ref{fig2}), H.\,P.\ Tranitz and W.\ Wegscheider (Fig.~\ref{fig3}) for providing the wafers, and J.\ Lapointe for assistance with nanofabrication (Fig.~\ref{fig2}). Financial support was provided by the DFG via SFB 631, NSERC, CIAR, the BMBF via DIP-H.2.1, and the German Excellence Initiative via the "Nanosystems Initiative Munich (NIM)".


\begin{thebibliography}{apssamp}
\bibitem{comm-1} An additional overall shift of the stability diagram is caused by the common voltage $V_{\rm QPC}$ applied to all ohmic contacts except I, where $I_{\rm QPC}$ is detected.
\bibitem{comment} Precise predictions of the limit placed on the coherence time anywhere in the stability diagram could be achieved with broad band counting statistics measurements.
\end{thebibliography}
\end{document}